\documentclass[dvips]{article}
\usepackage{epsfig}
\newcommand{\ie}{{\it i.e.},\,}
\newcommand{\etal}{{\it et al.\ }}

\newcommand{\zdot}{\makebox[0pt][l]{.}}
\newcommand{\up}[1]{\ifmmode^{\rm #1}\else$^{\rm #1}$\fi}

\newcommand{\uph}{\up{h}}
\newcommand{\upm}{\up{m}}
\newcommand{\ups}{\up{s}}
\newcommand{\arcd}{\ifmmode^{\circ}\else$^{\circ}$\fi}
\newcommand{\arcm}{\ifmmode{'}\else$'$\fi}
\newcommand{\arcs}{\ifmmode{''}\else$''$\fi}

\newcommand{\Abstract}[2]{{\footnotesize\begin{center}ABSTRACT\end{center}
\vspace{1mm}\par#1\par
\noindent
{~}{\it #2}}}

\newcommand{\TabCap}[2]{\begin{center}\parbox[t]{#1}{\begin{center}
  \small {\spaceskip 2pt plus 1pt minus 1pt T a b l e}
  \refstepcounter{table}\thetable \\[2mm]
  \footnotesize #2 \end{center}}\end{center}}

\newcommand{\FigCap}[1]{\footnotesize\par\noindent Fig.\  %
  \refstepcounter{figure}\thefigure. #1\par}

\newcommand{\TableFont}{\footnotesize}

\newcommand{\MakeTable}[4]{\begin{table}[htb]\TabCap{#2}{#3}
  \begin{center} \TableFont \begin{tabular}{#1} #4 
  \end{tabular}\end{center}\end{table}}

\newcommand{\MakeTableSep}[4]{\begin{table}[p]\TabCap{#2}{#3}
  \begin{center} \TableFont \begin{tabular}{#1} #4 
  \end{tabular}\end{center}\end{table}}

\newenvironment{references}%
{
\footnotesize \frenchspacing

\newcommand{\ApJ}{Astrophys.\ J.}

\newcommand{\Acta}{Acta Astron.}

\renewcommand{\and}{{\rm and }}
\section{{\rm REFERENCES}}
\sloppy \hyphenpenalty10000
\begin{list}{}{\leftmargin1cm\listparindent-1cm
\itemindent\listparindent\parsep0pt\itemsep0pt}}%
{\end{list}\vspace{2mm}}

\def\TYLDA{~}
\newlength{\DW}
\newcommand{\refitem}[5]{\item[]{#1} #2%
\def\REFARG{#3}\ifx\REFARG\TYLDA\else, {\it#3}\fi
\def\REFARG{#4}\ifx\REFARG\TYLDA\else, {\bf#4}\fi
\def\REFARG{#5}\ifx\REFARG\TYLDA\else, {#5}\fi.}

\newcommand{\Section}[1]{\section{#1}}

\newcommand{\Acknow}[1]{\par\vspace{5mm}{\bf Acknowledgements.} #1}
\newfont{\bb}{ptmbi8t at 12pt}

\newcommand{\uprule}{\rule{0pt}{2.5ex}}

\newcommand{\dorule}{\rule[-2ex]{0pt}{2ex}}

\begin{document}

\begin{center}
{\Large\bf Contact Binaries in OGLE-I Database\footnote{Based on observations 
obtained at the Las Campanas Observatory of the Carnegie Institution of 
Washington.}} 

\vspace{8mm}
{\bf M. ~S~z~y~m~a~\'n~s~k~i,~~ M. ~K~u~b~i~a~k~~ and~~ A. ~U~d~a~l~s~k~i}\\
Warsaw University Observatory, Al.~Ujazdowskie~4, 00-478~Warsaw, Poland\\ 
e-mail (msz,mk,udalski)@astrouw.edu.pl
\end{center}

\vspace{8mm}
\Abstract{We present a new catalog of 1575 contact binary stars fainter than 
${I=18}$~mag identified in OGLE-I database in the selected directions toward 
the Galactic bulge and the Galactic bar. For completeness we recall here also 
the objects from the formerly published OGLE catalog of contact binaries 
brighter than ${I=18}$~mag. 
 
To ease the comparison and/or distinction between contact binaries with 
sinusoidal light curves and  pulsating stars we also give a catalog of 506 
objects with nearly sinusoidal light curves which were classified or 
re-classified as pulsating variable stars.  

Present catalog contains the most numerous, observationally homogeneous sample 
of contact binaries. Given are the statistical properties of the sample 
relating to the distributions of brightness, periods and amplitudes of the 
contact binaries.}{}

\Section{ Introduction}
The Optical Gravitational Lensing Experiment (OGLE) is a long term project of 
searching for the dark matter in the Universe  with microlensing phenomena 
(Paczy\'nski 1986, Udalski \etal 1993b). The first phase of the project 
(OGLE-I) started in 1992 and lasted over four consecutive observing seasons  
till 1995. The 1-m Swope telescope at the Las Campanas Observatory in Chile, 
operated by the Carnegie Institution of Washington, was used for observations. 
During the entire program the detector was a ${2048\times2048}$ Ford/Loral CCD 
camera. Its pixel size, 15~$\mu$m, corresponded to the image  scale of 
0.44~arcsec/pixel -- sufficient for resolving stars in  dense  fields. 
Eighteen  fields ${15\times15}$~arcmin were monitored during a period spanning 
more than 1200 days providing more than 240 observations of each field 
through filter $I$. Each field was also observed, although less frequently 
(20--30 times), through filter $V$. The coordinates of the centers of the 
fields are given in Table~1. Fields with names beginning with the letters 
``BW'' are located in the Baade's window and those beginning with the letters 
``MM'' -- in the Galactic bar. The last column of Table~1 gives the total 
number of stars distinguished on the $I$-band template image of each field. 
\MakeTable{|l|c|c|r|c|c|}{6cm}{OGLE-I Fields}{
\hline
Field & RA (2000.0)  & DEC (2000.0) & \multicolumn{1}{|c|}{$l$} & $b$ & Number of stars\\
      &              &              &     &     &   on a template\\
\hline
BW1  & 18\uph02\upm24\ups& $-29\arcd49\arcm05\arcs$ & $ 1\zdot\arcd1$& $-3\zdot\arcd6$& 207251\\
BW2  & 18\uph02\upm24\ups& $-30\arcd15\arcm05\arcs$ & $ 0\zdot\arcd7$& $-3\zdot\arcd8$& 234443\\
BW3  & 18\uph04\upm24\ups& $-30\arcd15\arcm05\arcs$ & $ 0\zdot\arcd9$& $-4\zdot\arcd2$& 174349\\
BW4  & 18\uph04\upm24\ups& $-29\arcd49\arcm05\arcs$ & $ 1\zdot\arcd3$& $-4\zdot\arcd0$& 234258\\
BW5  & 18\uph02\upm24\ups& $-30\arcd02\arcm05\arcs$ & $ 0\zdot\arcd9$& $-3\zdot\arcd7$& 197246\\
BW6  & 18\uph03\upm24\ups& $-30\arcd15\arcm05\arcs$ & $ 0\zdot\arcd8$& $-4\zdot\arcd0$& 226301\\
BW7  & 18\uph04\upm24\ups& $-30\arcd02\arcm05\arcs$ & $ 1\zdot\arcd1$& $-4\zdot\arcd1$& 193559\\
BW8  & 18\uph03\upm24\ups& $-29\arcd49\arcm05\arcs$ & $ 1\zdot\arcd2$& $-3\zdot\arcd8$& 233591\\
BW9  & 18\uph00\upm50\ups& $-29\arcd49\arcm05\arcs$ & $ 0\zdot\arcd9$& $-3\zdot\arcd3$& 265840\\
BW10 & 18\uph00\upm50\ups& $-50\arcd02\arcm05\arcs$ & $ 0\zdot\arcd7$& $-3\zdot\arcd4$& 255248\\
BW11 & 18\uph00\upm50\ups& $-30\arcd15\arcm05\arcs$ & $ 0\zdot\arcd5$& $-3\zdot\arcd5$& 250170\\
BWC  & 18\uph03\upm24\ups& $-30\arcd02\arcm00\arcs$ & $ 1\zdot\arcd0$& $-3\zdot\arcd9$& 254481\\
MM1-A& 18\uph06\upm52\ups& $-26\arcd38\arcm05\arcs$ & $ 4\zdot\arcd4$& $-2\zdot\arcd9$& 242602\\
MM1-B& 18\uph06\upm52\ups& $-26\arcd51\arcm05\arcs$ & $ 4\zdot\arcd2$& $-3\zdot\arcd0$& 247554\\
MM5-A& 17\uph47\upm30\ups& $-34\arcd45\arcm00\arcs$ & $-4\zdot\arcd8$& $-3\zdot\arcd4$& 231279\\
MM5-B& 17\uph47\upm30\ups& $-34\arcd57\arcm00\arcs$ & $-4\zdot\arcd9$& $-3\zdot\arcd5$& 175449\\
MM7-A& 18\uph10\upm53\ups& $-25\arcd54\arcm20\arcs$ & $ 5\zdot\arcd4$& $-3\zdot\arcd3$& 221018\\
MM7-B& 18\uph11\upm47\ups& $-25\arcd54\arcm20\arcs$ & $ 5\zdot\arcd5$& $-3\zdot\arcd5$& 160630\\
\hline}

Observations were reduced practically on-line with the procedure described in 
detail in Udalski \etal (1992) and the results were collected in a database as  
described by Szyma\'nski and Udalski (1993a). Udalski \etal (1992) also gave 
a detailed analysis of photometric errors. It will be sufficient to recall 
here that the errors of single observation  depend on brightness and vary 
from about  0.015--0.020~mag (depending on the  image quality) at 
${I{=}14.5}$~mag to about 0.08--0.13~mag at ${I=18.5}$~mag. Unfortunately the 
magnitude scale is somewhat non-linear due to non-linearity of the CCD 
detector used. While the accuracy of standard system magnitudes is about ${\pm 
0.04}$~mag for the brighter stars, it drops to about ${\pm0.10}$~mag at the 
faint end. 

One of the ``by-products'' of the OGLE-I observations was the Catalog of 
Periodic Variable Stars in the Galactic bulge published in a series of papers 
by Udalski \etal (1994, 1995a,b, 1996, 1997). Due to a large number of 
observed stars the Catalog was limited to objects brighter than ${I=18.0}$~mag. 
The upper limit of brightness, ${I\approx14}$~mag, resulted from saturation of 
bright star images on the CCD detector. 

\Section{Contact Binaries in the OGLE-I Database}
The already published  Catalog of Periodic Variable Stars in the Galactic 
bulge lists, among others, 1166 objects classified as contact binaries (EW). 
The only criterion of classification was the shape of their light curves. In 
the case of relatively bright stars for which observational errors are small 
and the definition of the light curve is good, this was enough for unambiguous 
classification. 

We extended the search for variable stars to the objects with $I$ brightness 
between 18 and about 19.5~mag. Stars searched for variability were extracted 
from the {\it I}-band database in the same way as in the former Catalog, \ie 
the star should have at least 40 good observations (typically 100--200 
observations) and the standard deviation of the magnitude had to be larger 
than the sigma limit for non-variable stars of a given magnitude (Udalski 
\etal 1993a). The selected objects were subject to period search procedure 
both with the classic periodogram analysis and with the phase dispersion 
minimization method of Stellingwerf (1978). About two thousand new objects 
with periodic light variations were found. The majority of them had light 
curves typical for contact binaries and their identification was 
straightforward. The well known problem, however, posed the distinguishing 
between sinusoidal pulsators and contact binaries with sinusoidal shape of 
light curve and two times longer periods. Although this distinction may be 
very difficult in particular cases it can be done in a statistical sense. 

\begin{figure}[htb]
\vglue-7mm
\centerline{\includegraphics[width=11.5cm]{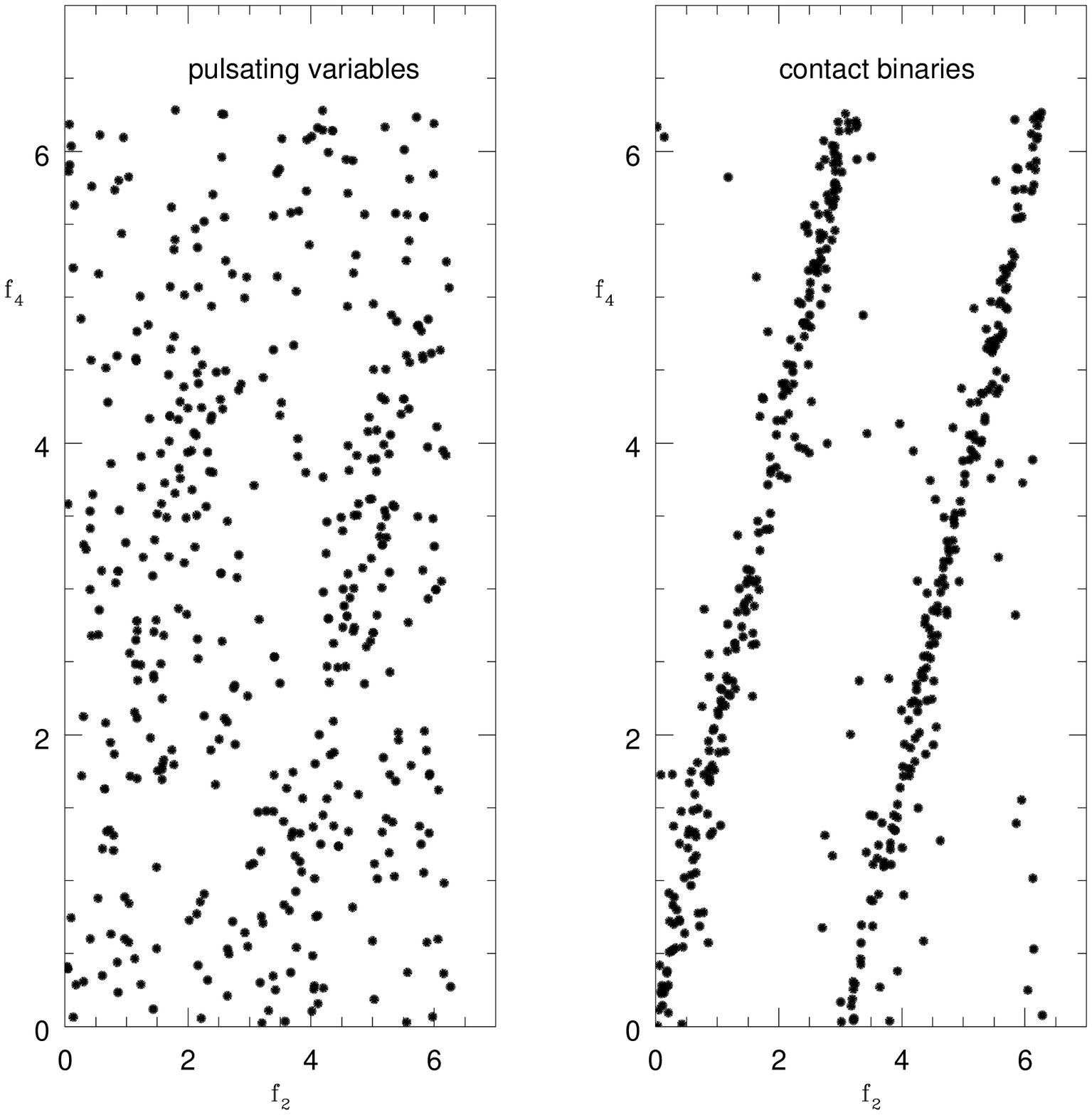}}
\vglue-3mm
\FigCap{Relation between phase constants $f_2$ and $f_4$ (from  Eq.~1) for 
pulsating stars (left panel) and contact binaries (right panel). The relation 
between $f_2$ and $f_4$ in the right panel is equivalent to ${f_2-2f_4=0}$. 
Phases are expressed in radians.}
\end{figure} 

To show this we used for  quantitative description of the light curves the 
Fourier decomposition method in which the observed light variations $A(t)$, 
with a period ${P=2\pi/\omega}$ known from periodogram analysis, are 
approximated by the sum of harmonics 
$$A(t)=a_0+\sum_i a_i\cos(i\cdot\omega t+f_{i})\eqno(1)$$

We performed these calculations up to the fifth harmonic for all objects with 
periodic light variations, adopting such values of periods that the light 
curves showed two minima in one cycle, \ie we treated all variable stars as 
EW-type objects. From them we selected two groups of about four hundred 
objects each: evident pulsators \ie objects  with periods shorter than about 
0.2~d and/or with shapes of light curves that could not be mistaken with 
EW-type variable stars and evident W~UMa objects (randomly selected bright 
objects with well defined EW-type light curves). The relations between phases 
$f_2$ and $f_4$ for both groups are shown in Fig.~1. It can easily be seen 
that there exists a well defined relation between these two phases for contact 
binaries, whereas the same relations for pulsating variable stars is very 
weak. Using this result in doubtful cases, when the unambiguous identification 
of the light curve was not obvious, the objects fulfilling the relation for 
contact binaries were classified as EW and the objects laying off this 
relation -- as pulsating variables P. We believe that this procedure (applied 
to less than  10\% of all  objects with equal depths of minima which appeared 
to be "doubtful") reduces the expected number of misidentifications in our 
samples of EW and pulsating variable stars down to  about one  percent. 

One of the easily recognized features of the contact binary type  light curve 
is the relative depth of eclipses. By visual inspection we divided our sample 
of EW variable stars into two groups: ``EWa'' with markedly different depths 
of minima,  and ``EWs'' -- for which, within the error limits, the depths of 
both minima are the same. The two groups are statistically  and probably also 
physically nonequivalent. The EWs group contains objects with the same or very 
close surface temperature of the components, whereas the different depths of 
eclipses suggest different temperatures. On the other hand, the sample of 
stars with different depths of minima is more uniform in the sense that it is 
not contaminated with single pulsating stars: contact binaries with different 
depths of minima can hardly be mistaken with pulsating variable stars. 

\MakeTableSep{|l|r|c|c|cc|c|c|}{12cm}{A sample page from the the table
``EWa.new'' containing newly identified contact binaries in OGLE-I
database.}{
\hline
\uprule
Field & \multicolumn{1}{|c|}{No} & RA(2000.0) & DEC(2000.0) & $I$   & $V-I$ & Period & Type\\ 
      &    &            &             & [mag] & [mag] & [d]    & \dorule\\
\hline
\uprule
BW9   &  189794  & 18\uph00\upm49\zdot\ups45 & $-29\arcd41\arcm53\zdot\arcs6$ &  18.83&   2.22&    .20680  &EWa   \\
MM1-A &  128388  & 18\uph06\upm52\zdot\ups74 & $-26\arcd35\arcm41\zdot\arcs6$ &  18.98&   1.60&    .23516  &EWa   \\
BW5   &  166779  & 18\uph02\upm35\zdot\ups99 & $-29\arcd55\arcm37\zdot\arcs1$ &  18.58&   9.99&    .23919  &EWa   \\
MM1-A &  113202  & 18\uph06\upm48\zdot\ups63 & $-26\arcd42\arcm03\zdot\arcs4$ &  18.05&   2.02&    .24168  &EWa   \\
MM1-B &  163429  & 18\uph06\upm54\zdot\ups64 & $-26\arcd49\arcm18\zdot\arcs9$ &  18.42&   1.77&    .24442  &EWa   \\
MM1-A &   11980  & 18\uph06\upm18\zdot\ups58 & $-26\arcd39\arcm47\zdot\arcs4$ &  18.12&   1.78&    .25362  &EWa   \\
BW3   &  168639  & 18\uph04\upm45\zdot\ups11 & $-30\arcd11\arcm40\zdot\arcs5$ &  18.18&   1.91&    .25980  &EWa   \\
BW10  &    3060  & 18\uph00\upm18\zdot\ups98 & $-30\arcd07\arcm52\zdot\arcs9$ &  18.53&   1.66&    .26218  &EWa   \\
MM1-B &   28221  & 18\uph06\upm15\zdot\ups34 & $-26\arcd45\arcm39\zdot\arcs9$ &  17.63&   1.86&    .26446  &EWa   \\
BW8   &   70608  & 18\uph03\upm07\zdot\ups78 & $-29\arcd54\arcm13\zdot\arcs5$ &  17.84&   1.63&    .26598  &EWa   \\
MM7-B &  108382  & 18\uph11\upm50\zdot\ups75 & $-25\arcd52\arcm39\zdot\arcs8$ &  18.53&   2.04&    .26708  &EWa   \\
MM7-A &   65167  & 18\uph10\upm43\zdot\ups67 & $-26\arcd00\arcm44\zdot\arcs4$ &  19.73&   9.99&    .26900  &EWa   \\
BW1   &  174918  & 18\uph02\upm37\zdot\ups68 & $-29\arcd43\arcm44\zdot\arcs8$ &  19.12&   9.99&    .26986  &EWa   \\
BW4   &  148730  & 18\uph04\upm26\zdot\ups32 & $-29\arcd48\arcm48\zdot\arcs6$ &  18.67&   1.57&    .27030  &EWa   \\
BW6   &  139978  & 18\uph03\upm29\zdot\ups76 & $-30\arcd17\arcm19\zdot\arcs2$ &  18.44&   1.76&    .27304  &EWa   \\
MM1-B &   58328  & 18\uph06\upm33\zdot\ups60 & $-26\arcd47\arcm27\zdot\arcs1$ &  17.23&   1.53&    .27900  &EWa   \\
BW2   &   63876  & 18\uph02\upm09\zdot\ups37 & $-30\arcd22\arcm34\zdot\arcs8$ &  18.18&   2.05&    .28142  &EWa   \\
BW11  &   52517  & 18\uph00\upm24\zdot\ups28 & $-30\arcd14\arcm45\zdot\arcs3$ &  19.42&   9.99&    .28434  &EWa   \\
BW9   &   62576  & 18\uph00\upm28\zdot\ups34 & $-29\arcd47\arcm29\zdot\arcs7$ &  19.51&   2.14&    .28612  &EWa   \\
BW11  &   75418  & 18\uph00\upm39\zdot\ups94 & $-30\arcd19\arcm59\zdot\arcs0$ &  18.62&   1.48&    .28780  &EWa   \\
BW7   &   16970  & 18\uph03\upm53\zdot\ups37 & $-30\arcd00\arcm02\zdot\arcs4$ &  18.88&   1.50&    .28838  &EWa   \\
BW3   &   78973  & 18\uph04\upm23\zdot\ups74 & $-30\arcd21\arcm41\zdot\arcs1$ &  19.01&   9.99&    .28998  &EWa   \\
BW2   &   65073  & 18\uph02\upm08\zdot\ups60 & $-30\arcd21\arcm51\zdot\arcs3$ &  18.98&   1.59&    .29032  &EWa   \\
BW9   &  147608  & 18\uph00\upm43\zdot\ups74 & $-29\arcd45\arcm11\zdot\arcs4$ &  19.15&   1.69&    .29040  &EWa   \\
BW9   &   43248  & 18\uph00\upm27\zdot\ups40 & $-29\arcd53\arcm33\zdot\arcs6$ &  18.92&   1.81&    .29102  &EWa   \\
BW8   &  199379  & 18\uph03\upm37\zdot\ups68 & $-29\arcd42\arcm28\zdot\arcs8$ &  18.45&   1.72&    .29134  &EWa   \\
BW7   &  132331  & 18\uph04\upm26\zdot\ups34 & $-29\arcd57\arcm55\zdot\arcs7$ &  18.85&   1.66&    .29176  &EWa   \\
MM1-A &  130188  & 18\uph06\upm45\zdot\ups13 & $-26\arcd34\arcm59\zdot\arcs0$ &  19.23&   1.71&    .29238  &EWa   \\
BW10  &   78472  & 18\uph00\upm36\zdot\ups63 & $-30\arcd06\arcm30\zdot\arcs9$ &  17.93&   1.51&    .29456  &EWa   \\
BWc   &    1362  & 18\uph02\upm51\zdot\ups58 & $-30\arcd09\arcm19\zdot\arcs9$ &  18.86&   1.37&    .29694  &EWa   \\
MM5-B &  169652  & 17\uph47\upm48\zdot\ups61 & $-34\arcd54\arcm01\zdot\arcs8$ &  18.05&   1.54&    .29770  &EWa   \\
BW2   &   25180  & 18\uph01\upm52\zdot\ups88 & $-30\arcd11\arcm31\zdot\arcs8$ &  19.54&   1.14&    .29840  &EWa   \\
BWc   &  205530  & 18\uph03\upm38\zdot\ups37 & $-30\arcd00\arcm29\zdot\arcs2$ &  19.02&   1.61&    .29860  &EWa   \\
MM1-A &  234696  & 18\uph07\upm12\zdot\ups70 & $-26\arcd33\arcm17\zdot\arcs4$ &  18.47&   1.90&    .29872  &EWa   \\
BW10  &   56593  & 18\uph00\upm28\zdot\ups33 & $-30\arcd00\arcm19\zdot\arcs8$ &  18.28&   2.35&    .29880  &EWa   \\
MM1-B &   79232  & 18\uph06\upm36\zdot\ups89 & $-26\arcd54\arcm09\zdot\arcs7$ &  18.31&   1.84&    .29906  &EWa   \\
MM1-A &   42504  & 18\uph06\upm35\zdot\ups91 & $-26\arcd42\arcm10\zdot\arcs9$ &  19.46&   1.46&    .29938  &EWa   \\
MM1-A &  197164  & 18\uph07\upm10\zdot\ups15 & $-26\arcd35\arcm16\zdot\arcs4$ &  18.69&   9.99&    .30160  &EWa   \\
BW9   &  178667  & 18\uph00\upm52\zdot\ups53 & $-29\arcd47\arcm12\zdot\arcs5$ &  18.49&   1.38&    .30186  &EWa   \\
MM7-A &  184242  & 18\uph11\upm06\zdot\ups08 & $-25\arcd48\arcm42\zdot\arcs8$ &  19.44&   9.99&    .30228  &EWa   \\
MM7-B &  115327  & 18\uph11\upm45\zdot\ups80 & $-25\arcd48\arcm25\zdot\arcs0$ &  18.01&   1.85&    .30252  &EWa   \\
BW2   &   20210  & 18\uph01\upm50\zdot\ups69 & $-30\arcd14\arcm25\zdot\arcs5$ &  19.40&   9.99&    .30260  &EWa   \\
\hline}

\Section{The Catalog}
The catalog of  contact binaries can be found in anonymous FTP archive at

\centerline{\it ftp://ftp.astrouw.edu.pl/ogle/ogle1/contact\_binaries}
or on its US mirror at

\centerline{\it ftp://bulge.princeton.edu/ogle/ogle1/contact\_binaries}

The newly found W~UMa variable stars are listed in tables named EWa.new (595 
entries) and EWs.new (980 entries). For completeness, in tables EWa.old (525 
entries) and EWs.old (641 entries), we repeat here the contact binaries from 
the already published  Catalog of OGLE-I variable stars. The consecutive 
columns of these tables contain: (1) -- name of the OGLE-I field, (2) -- 
number of the object in OGLE-I database (3) -- Right Ascension (2000.0), (4) 
-- Declination (2000.0), (5) -- observed mean magnitude in filter $I$, (6) --  
observed ${V-I}$ (9.99 means lack of $V-I$ measurements), (7) --  period in 
days, (8) --  character of the light curve. The objects are ordered according 
to the increasing values of periods.  For illustration, the beginning of the 
table EWa.new is reproduced here in Table~2. 

The archive  contains also results of photometric observations of the stars 
from the catalog and finding charts. Please see README file there for detailed 
description of the archive contents. 

\begin{figure}[htb]
\vglue-7mm
\centerline{\includegraphics[width=11.3cm]{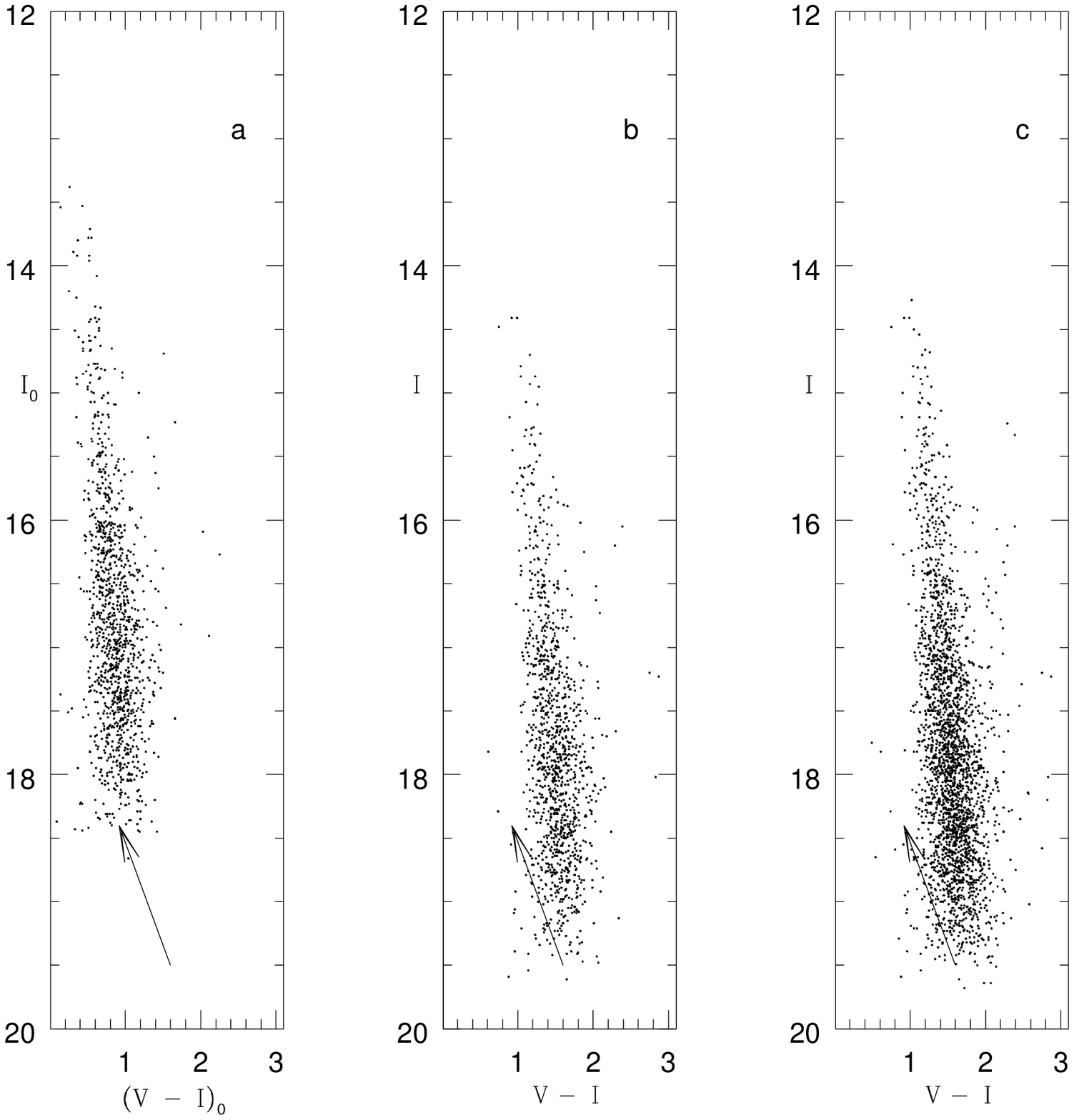}}
\FigCap{Color -- Brightness  diagram for contact binaries. (a) -- binaries 
corrected for interstellar extinction, (b) -- the same stars not corrected for 
interstellar extinction. (c) -- all binaries from our catalog. The arrow shows 
the average shift of both diagrams resulting from the 
absorption ${A_I=1.1}$~mag and ${E(V-I)=0.68}$~mag.}
\end{figure}

\Section{Color -- Brightness Diagram and Brightness Distribution of the 
Galactic Bulge Contact Binaries} 
Objects in our database are subject to interstellar extinction and reddening.  
Wo\'zniak and Stanek (1996) proposed an original method to investigate 
interstellar extinction based on two band photometry of the red clump stars. 
The method was applied by Stanek (1996) to the Baade's window in which the 
red clump stars are very abundant. He constructed an extinction map covering 
large part of the OGLE-I fields. We used this map to apply  the  extinction 
corrections to the observed $I$ and ${V-I}$ for the total of 1277 (new and 
old) objects located in the central part of Baade's window, covered by the 
Stanek's extinction map. The extinction-corrected color--magnitude diagram 
for these objects is given in panel (a) of Fig.~2. Panel (b) shows this 
relation for the same stars not corrected for extinction. Relation for all 
observed stars, not corrected for interstellar  extinction and reddening, is 
shown in panel (c). All diagrams are very similar; only the scatter of points   
in panel (a) is slightly smaller than in panel (b), suggesting that the 
applied extinction correction is physical and we are dealing, in major part at 
least, with objects located at the Galactic bulge distance. Were this not true 
then the application of the bulge value of extinction to our stars would be
meaningless. In any case we may state that the interstellar extinction does 
not influence markedly the statistical relation between colors and magnitudes 
of stars in our sample. This relation seems to be determined by the true 
relation for double stars, by the interstellar extinction, and the 
observational errors. Please also note that the relation in Fig.~2 differs 
from  the "main sequence" in the CMD diagrams given by Udalski \etal (1993a) 
for all stars observed in the direction of the Baade's window and usually 
connected with the Galactic disk stars. 

\begin{figure}[htb]
\vglue-7mm
\centerline{\includegraphics[width=11.8cm]{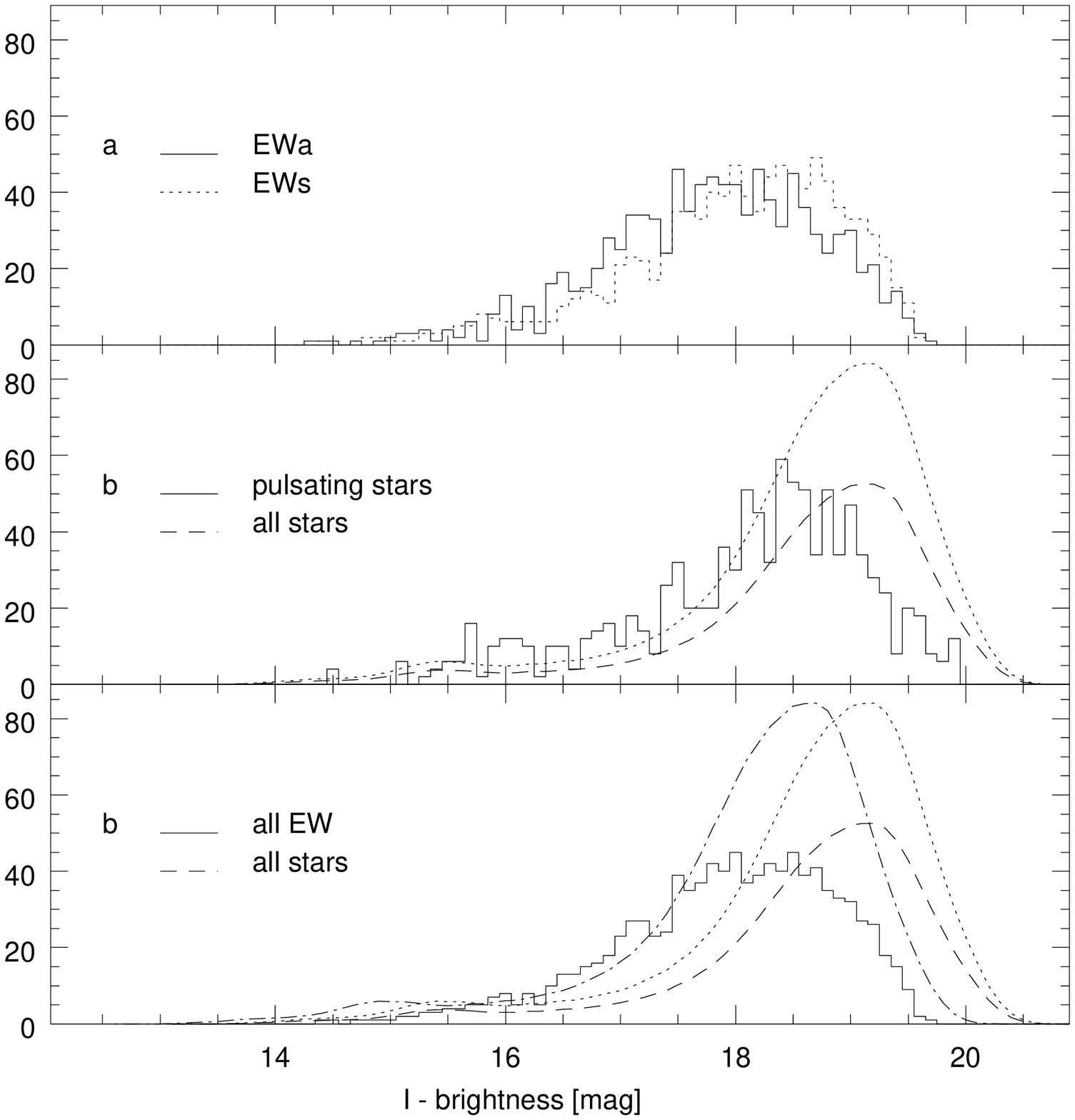}}
\FigCap{Distribution of the observed {\it I}-band brightness of contact 
binaries from the present catalog: (a) -- all EWa (1119) and all EWs (1621) 
binaries, not corrected for the interstellar extinction. (b) -- histogram of 
brightness distribution of objects classified in our catalog as pulsating 
variables. Dashed line shows the brightness distribution of all stars 
identified in all fields in the direction of the Baade's window. Dotted line 
is the latter distribution multiplied by 1.6 to give the best fit to the 
histogram. (c) -- histogram of brightness distribution of all EW binaries 
compared with the distribution of brightness of all stars identified in all 
field in the direction of Baade's window (dashed curve). Dashed-dotted curve 
shows the latter distribution multiplied by 1.6 and shifted in $I$ scale by 
${-0.5}$~mag for the best fit with the histogram. The bin of the histograms  
is 0.1 mag; all  distributions are normalized to 1000 objects.} 
\end{figure}

The distribution of the observed {\it I}-band brightness of the contact 
binaries from the present catalog is shown in  Fig.~3. In panel (a) we compare 
these distributions for EWa and EWs objects. Except for the small, but 
apparently real, shift of EWs distribution toward fainter stars, both 
histograms look very similar and do not differ from the distribution of 
brightness of all EW binaries shown by the histogram in panel (c). The dashed  
curve in  panel (c) shows the observed brightness distribution of all stars 
identified in all fields in the  Baade's window. There are two reasons why 
these two distributions are different. Firstly, the brightness of a contact 
binary is greater than the brightness of similar single star, and secondly, 
because both distributions are normalized to the same number of objects, the 
apparent lack of faint contact binaries due to all possible selection effects 
results in corresponding rise of relative number of brighter objects. To 
account in approximate way for this effects we show in panel (b) of Fig.~3 the 
observed brightness distribution of pulsating stars from our sample and 
compare it with the brightness distribution of all stars. Both distributions 
are different for the same reasons as above, except for the fact that the mean 
brightness of the single pulsating star is the same as the brightness of 
constant star with the same physical characteristics. We also assume --  what 
may not be true -- that the efficiency of discovery of bright pulsating  
stars is essentially the same as for constant stars. The latter was estimated 
by Udalski \etal (1993a). Their conclusion was that this quantity is 
approximately constant (although less than 1) over the range of magnitudes 
between 15 and 18 in $I$ and diminishes rapidly for fainter stars. 

If we accept the above assumption  we may express the distributions from panel 
(b) of Fig.~3 as ${N_1\cdot f_1(I)}$ for constant stars and ${N_2\cdot 
f_2(I)a(I)}$ for pulsating stars. $N_1$ and $N_2$ are normalization factors, 
functions $f_1(I)$ and $f_2(I)$ describe true brightness distributions of all 
stars and pulsating stars, respectively; function $a(I)$ describes the 
detection probability of pulsation in stars of a given {\it I}-band brightness 
(irrespectively of the probability of identification of the object as a star, 
which is included in functions $f(I)$). 

If we assume additionally that pulsating stars are represented in the same 
proportion among the stars of all brightness, \ie ${f_1(I)=f_2(I)}$, and that 
for bright stars in {\it I}-band range 15--18~mag ${a(I)\approx1}$ then the 
distributions in this range are simply $N_1f(I)$ and $N_2f(I)$, and with our 
assumptions  should be the same. As it  can be seen from panel (b) in Fig.~3  
an acceptable agreement of both distributions for the stars brighter than 
about 18.5 mag  can be obtained by multiplying the star brightness 
distribution by 1.6 (dotted line). It means that due to the influence of the 
``filter'' $a(I)$, effective for the objects fainter than about 18~mag, the 
relative number of detected variable stars is reduced by a factor of 1/1.6. 

The comparison of the distribution multiplied by 1.6 (dotted line) with the 
histogram of brightness of contact binaries is shown in panel (c) of Fig.~3. 
The still remaining excess of contact binaries among bright stars is a result 
of their duplicity. Shift of about ${-0.5}$~mag in the $I$ scale seems to 
account for this effect (dashed-dotted line). 

The fair agreement of the histogram with the dashed-dotted curve on their 
rising branches suggests that our efficiency of discovering variable stars in 
$I$ brightness range 14--17.5~mag is the same as the efficiency of identifying 
constant stars and in this respect our sample is representative for the 
considered population of stars. The marked difference of both curves for 
fainter objects is a joint effect of the possible true difference in 
brightness distributions and the diminishing efficiency of variable star 
discovery. Separation of both effects does not seem to be possible. We can 
only draw a conclusion that the percentage of contact binary systems 
identified among the observed stars is smaller than the percentage of 
discovered pulsating stars and that the efficiency of discovering variable 
objects is not greater than  about 60\%. 

\Section{Period Distribution}
Period observations are practically free from selection effects that could 
affect directly the determined period value or the probability of the 
discovery of a variable object with a given period among the objects in a 
given range of brightness. The possible exception may be periods close to 
0.5~day or 1~day, difficult to be determined observationally. In general, 
however, the  period distribution is an important and observationally unbiased 
characteristic of a particular group of variable stars.

\begin{figure}[htb]
\vglue-7mm
\centerline{\includegraphics[width=10cm]{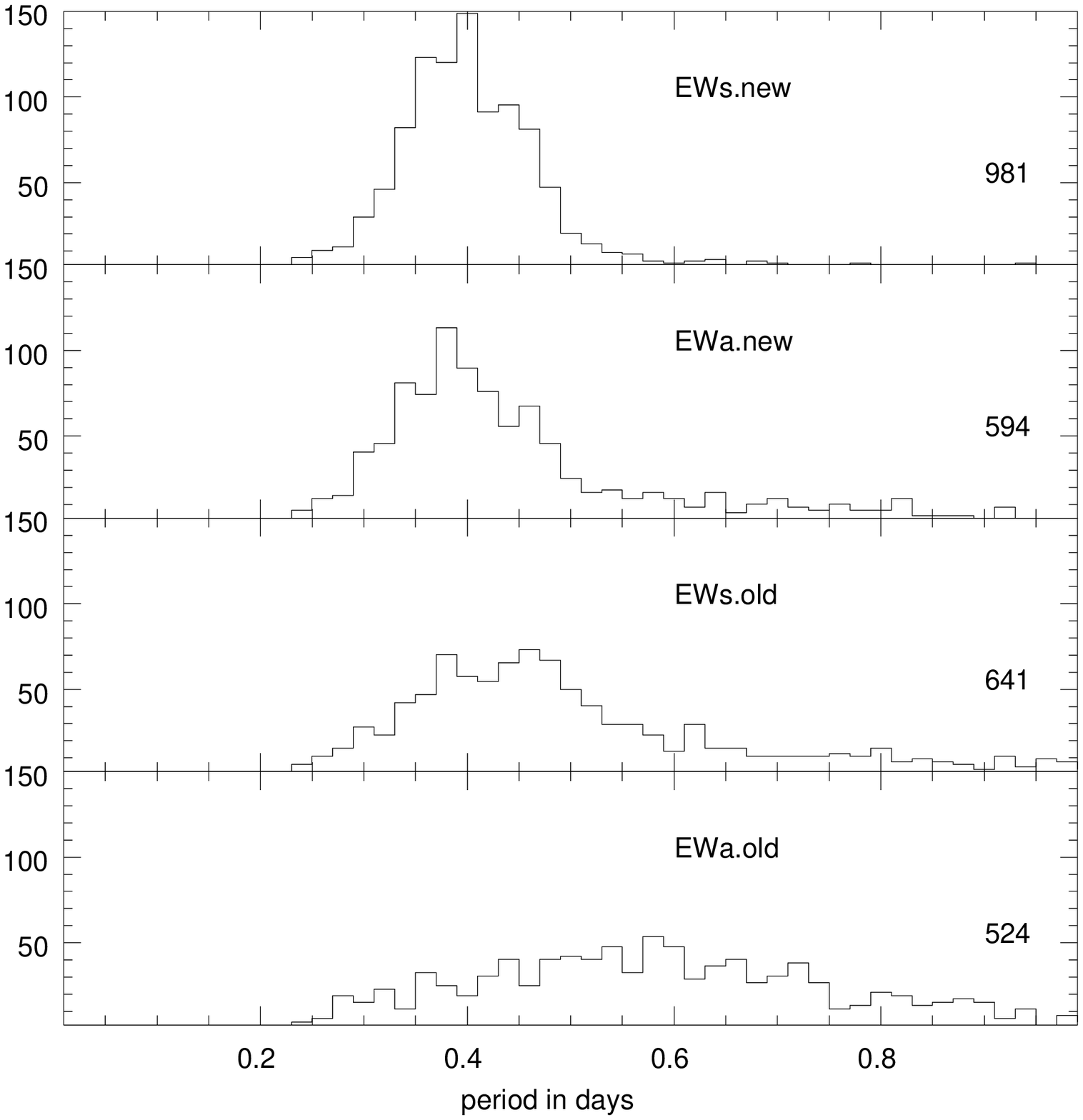}}
\FigCap{Period distributions of the different groups of  contact binaries in 
the present catalog. Period bins are 0.02~d wide. All distributions are 
normalized to 1000 objects; the real number of objects in each group is given 
in lower right corner of all panels.}
\end{figure} 
\begin{figure}[p]
\vglue-7mm
\centerline{\includegraphics[width=9.5cm]{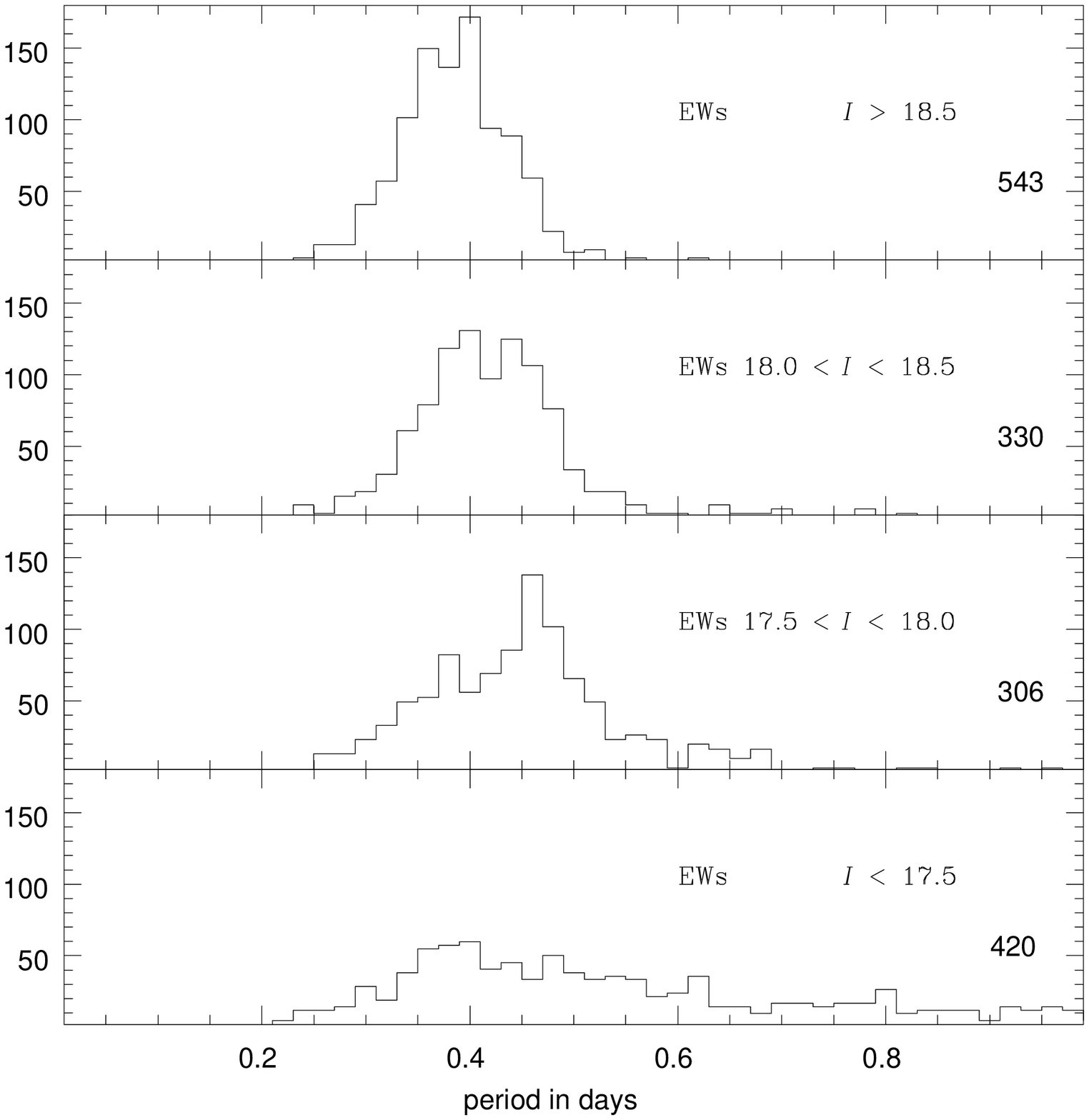}}
\FigCap{Period distributions for contact binaries with equal depths of minima 
in different ranges of brightness.}
\centerline{\includegraphics[width=9.5cm]{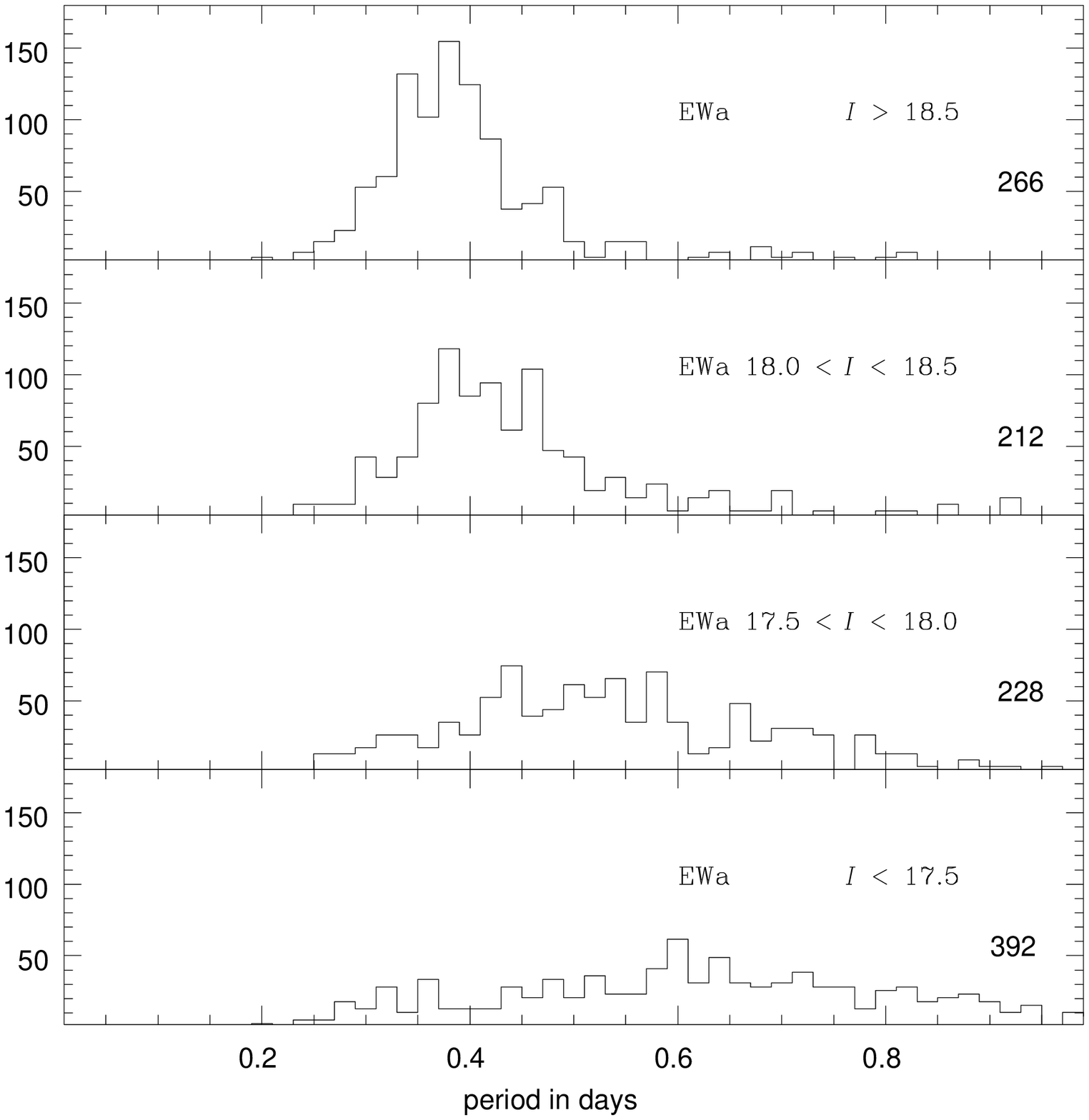}}
\FigCap{Period distributions for contact binaries with unequal depths of 
minima in different ranges of brightness.}
\end{figure} 
Fig.~4 shows the period distributions for contact binaries from the OGLE 
Catalog of variable stars and for the newly discovered objects. Histograms 
represent numbers of objects within period bins of 0.02~day. All the 
distributions are normalized to the same number (1000) of objects. The 
distributions are markedly different and we think it could be instructive to 
present them in more detail. 

Figs.~5 and 6 show period distributions for both groups of contact binaries 
(\ie with equal and unequal depths of minima) in different ranges of 
{\it I}-band brightness. It can be seen from Figs.~5 and 6 that the period 
distributions of EWa and EWs objects which are broad and different for stars 
brighter than ${I=18}$~mag become narrower and more similar for fainter stars. 
This reflects the character of the period -- brightness relation (see below). 

\begin{figure}[p]
\vglue-7mm
\centerline{\includegraphics[width=9.5cm]{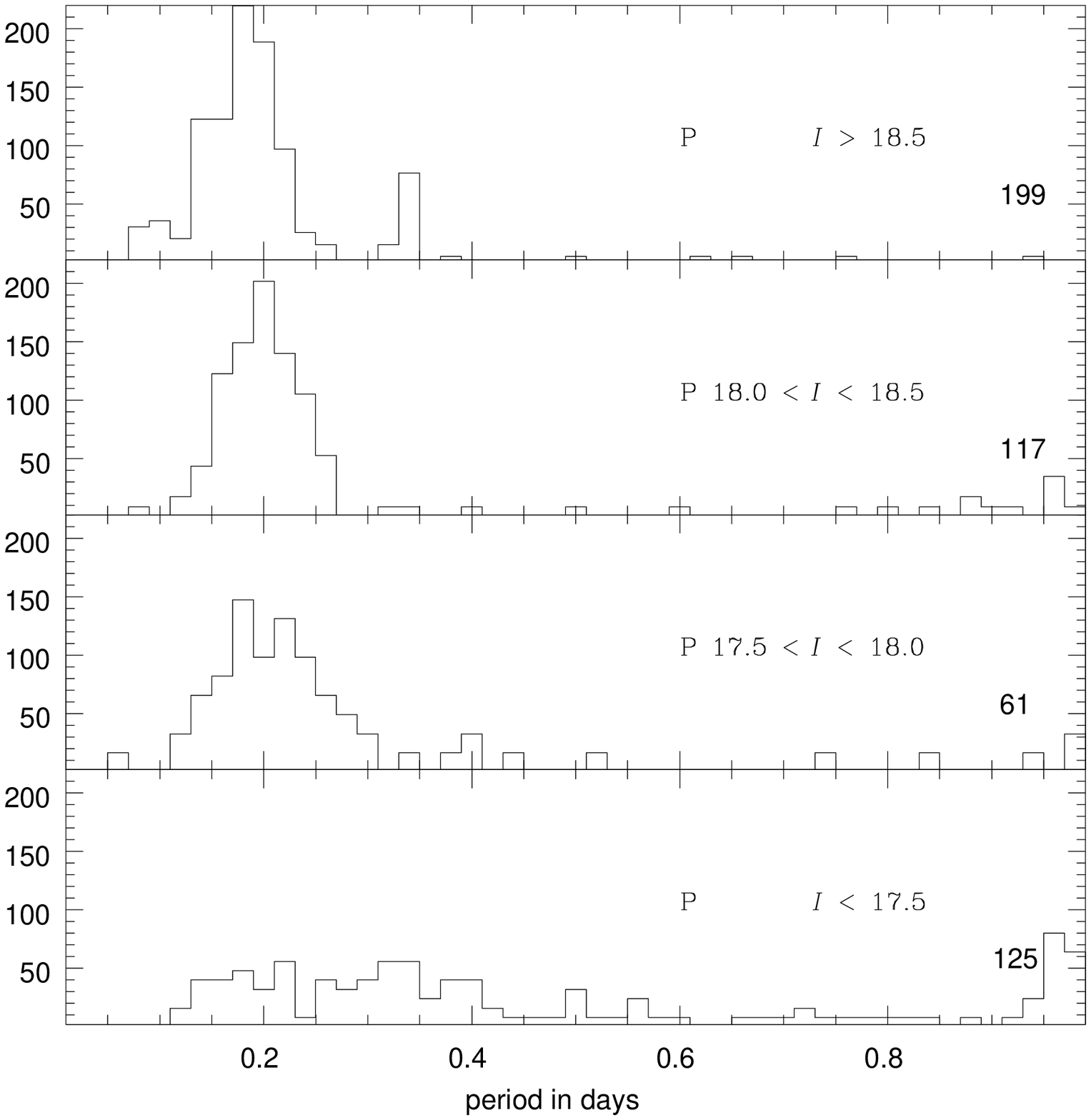}}
\vglue-2mm
\FigCap{Period distributions of the pulsating stars in the catalog for 
different ranges of brightness.}
\centerline{\includegraphics[width=9.5cm]{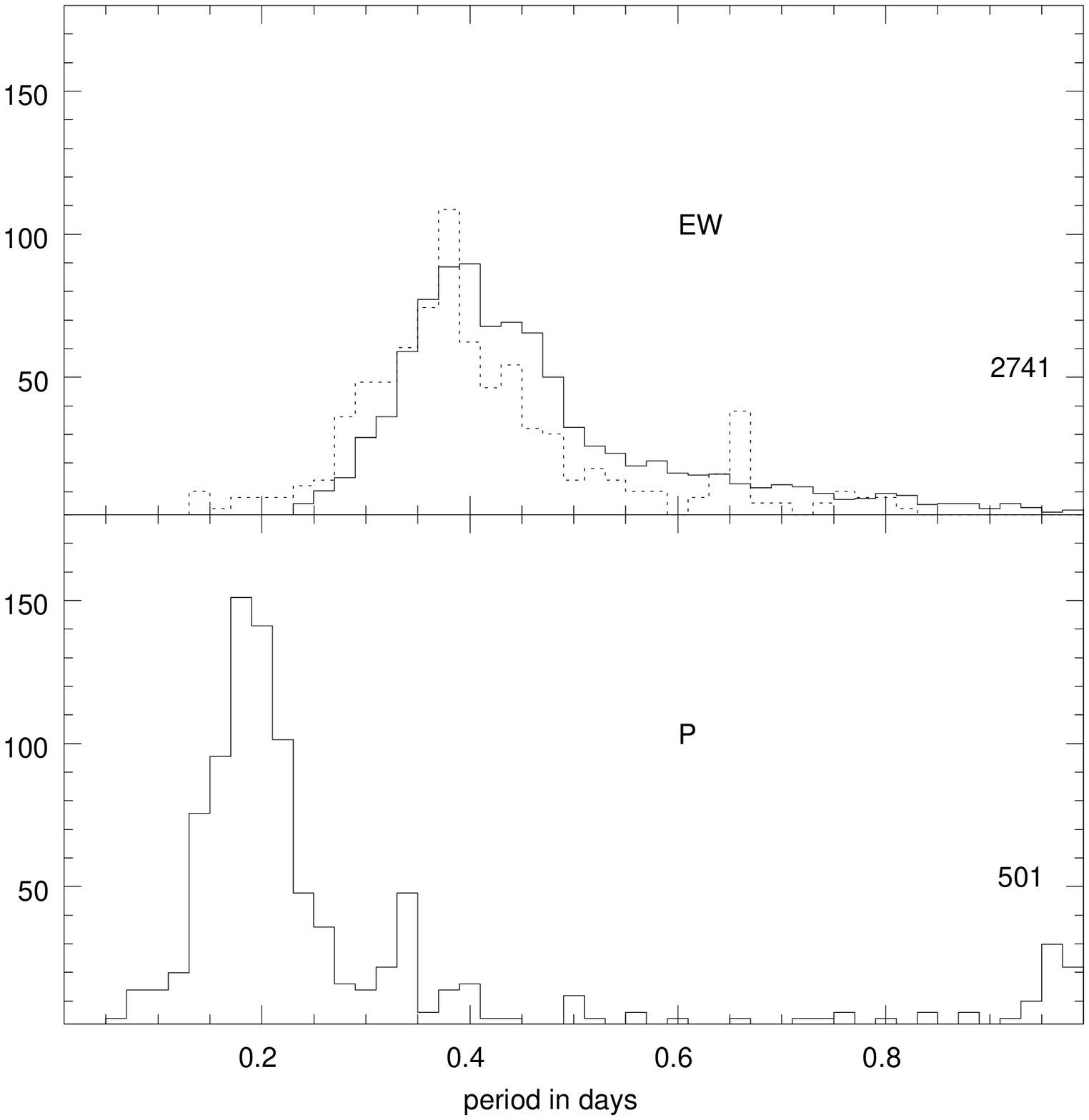}}
\vglue-2mm
\FigCap{Comparison of the  period distribution of all contact binaries (upper 
panel, continuous line) with that of the pulsating variables (lower panel). 
Dotted line in the upper panel shows the period distribution of pulsating 
stars obtained with the assumption that all of them are misclassified contact 
binaries with  periods two times longer. Histograms are normalized to 1000 
objects; the true numbers of EW and P objects are given in lower right 
corners of both panels.} 
\end{figure}
\begin{figure}[htb]
\vglue-7mm
\centerline{\includegraphics[width=9.5cm]{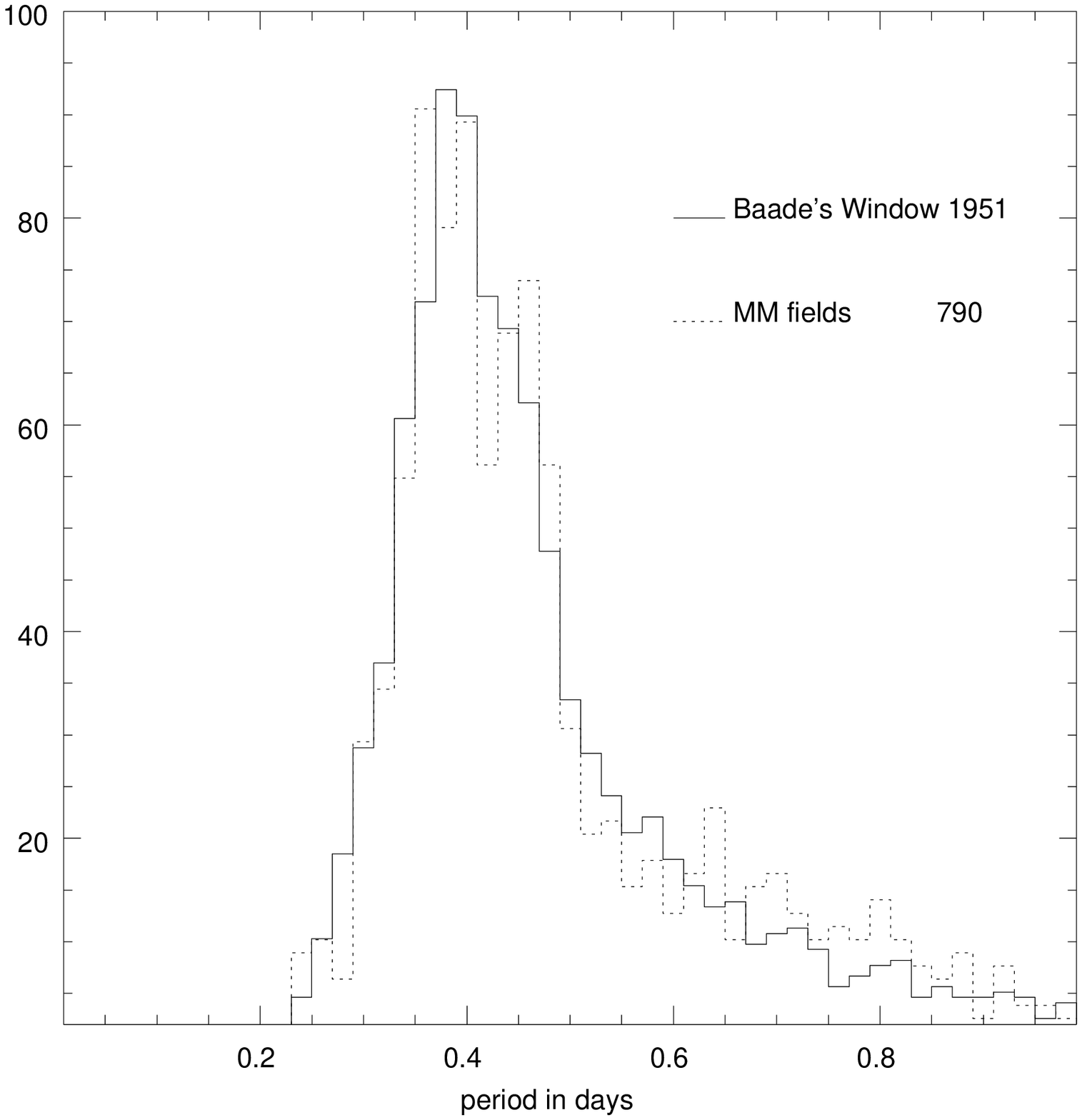}}
\FigCap{Comparison of period distributions of contact binaries from the 
central part of the  Baade's window (continuous line) and from the MM fields 
in the Galactic bar (dotted line).}
\end{figure} 

To check for the possible distortion of these distributions by misclassified 
pulsating stars we look closer at the period distributions of the stars 
classified as pulsating variables. They are shown in Fig.~7 for the same 
brightness ranges as before. Except for the brightest stars, ${I<17.5}$~mag, 
all the distributions seem to be concentrated around the period 0.2~d with no 
obvious dependence of the maximum position on brightness. Fig.~8 compares the 
period distributions of all EW binaries (upper panel, continuous line) and 
pulsating variables (lower panel). Distributions are normalized to 1000 
objects; true number of objects is also given in Fig.~8. Dotted line in the 
upper panel of Fig.~8 represents the period distribution of the pulsating 
stars from the lower panel obtained with the assumption that all of them are 
misclassified contact binaries with periods two times longer than pulsation 
periods. Both distributions look very  similar suggesting that erroneous 
classification of pulsating stars as contact binaries affects the period 
distribution of contact binaries uniformly in the whole range of periods, 
maybe with exception of the longest periods. The small shift of the  dotted 
histogram toward the smaller period values reflects the existence of very 
short periods among pulsating stars and their absence among contact binaries. 
Thus, the shape of the period distribution itself does not give information on 
the ``purity'' of our sample of contact binaries. We think, however, that the 
distinction between pulsating stars and contact binaries according to Fig.~1 
is statistically acceptable. 

During the first phase of the OGLE project we observed twelve field near the 
center of the  Baade's window and six fields in the Galactic bar  (``MM'' 
fields in Table~1). 790 contact binaries were identified in ``MM'' fields. 
Fig.~9 compares the distribution of their periods with the period distribution 
of contact binaries in the central part of the Baade's window. Within the 
scatter limit there is no difference between these two distributions. 

\Section{Period--Brightness Relation}
\begin{figure}[htb]
\vglue-7mm
\centerline{\includegraphics[width=11.8cm]{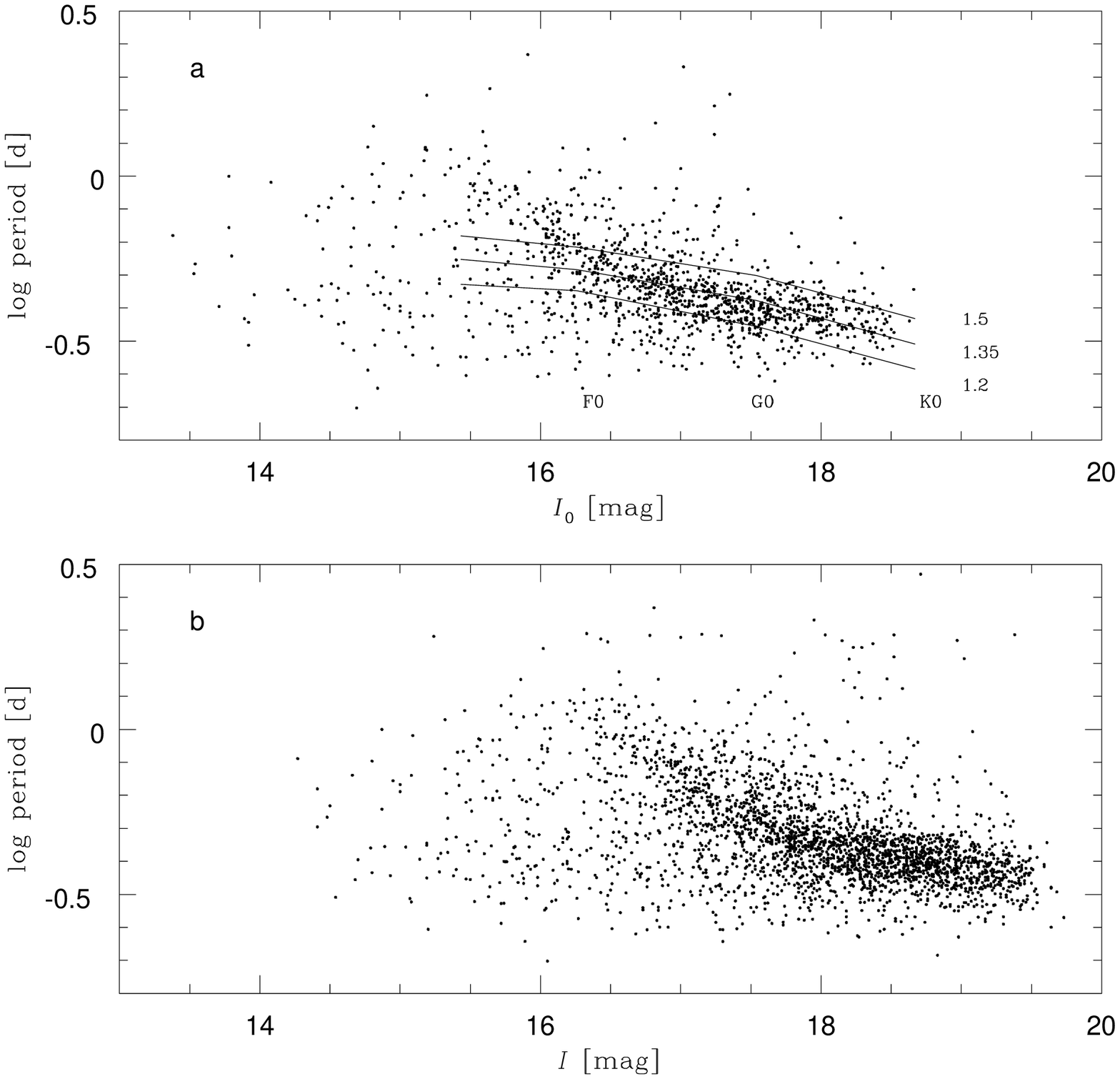}}
\FigCap{Period -- brightness relation for contact binaries in the present 
catalog. (a) -- objects corrected for the interstellar extinction. Continuous 
lines represent the expected period -- luminosity relation (shifted to fit 
the stars at the Galactic bulge distance) for binaries  composed of two equal 
main sequence stars in an orbit with semi axis equal to 1.2, 1.35, and 1.5 of 
their radii. (b) --the same relation for all stars not corrected for the 
interstellar extinction.}
\end{figure} 
\begin{figure}[htb]
\vglue-7mm
\centerline{\includegraphics[width=11.8cm]{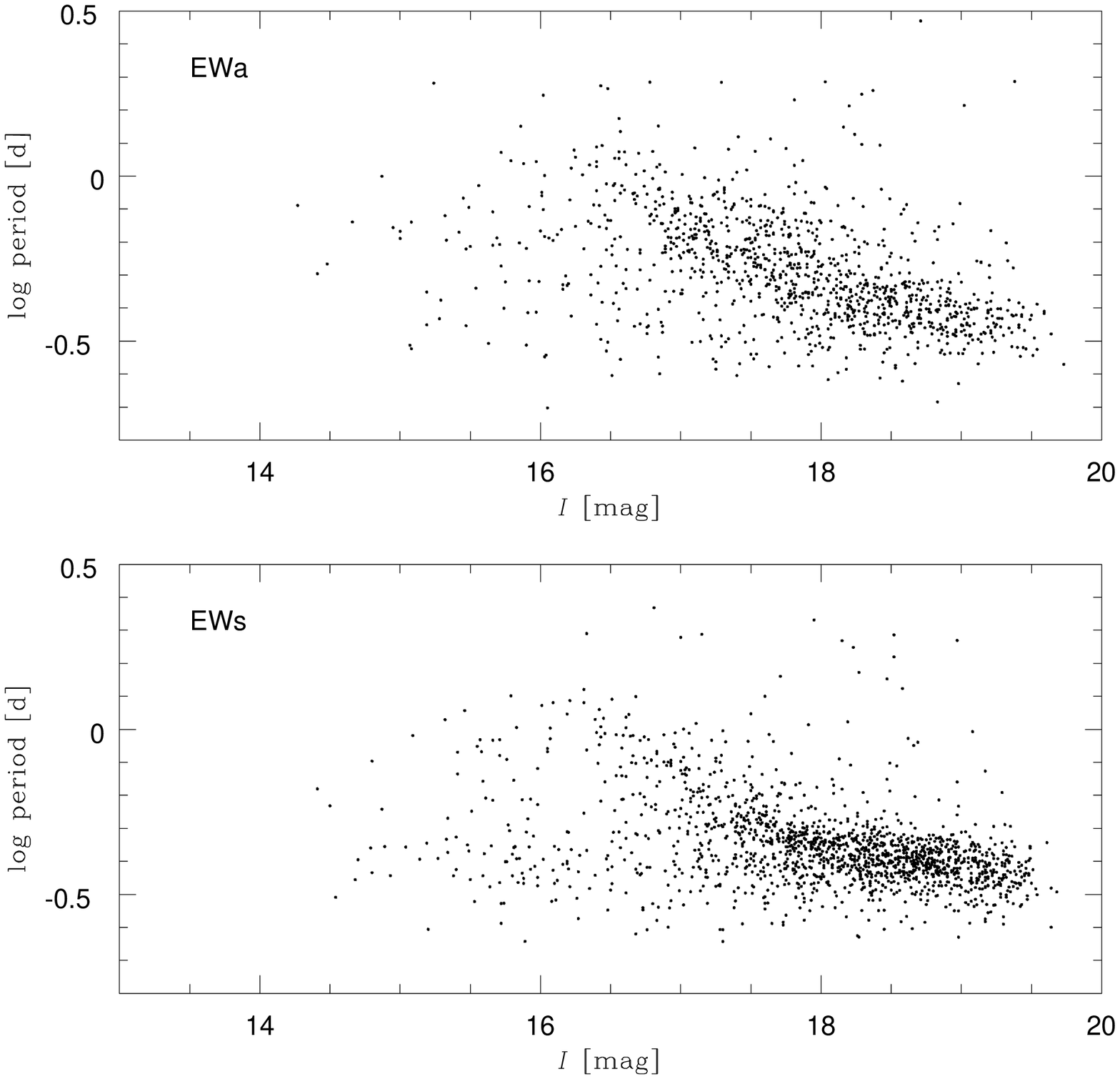}}
\FigCap{Period -- brightness relation for contact binaries from the present 
catalog, not corrected for the interstellar extinction. (a) -- stars with 
unequal depths of minima. (b) -- stars with equal depths of minima.} 
\end{figure} 

Construction  of the period--luminosity relation requires a knowledge of 
absolute magnitudes which are not known for our sample. We think, however, 
that some general features of the relation period -- observed brightness 
suggests that we are dealing with objects the majority of which is probably 
located at the distance of the Galactic bulge. Fig.~10 shows this relation for 
contact binaries from our sample: panel (a) -- for stars corrected for 
interstellar extinction, as described in Section~4, and panel (b) -- for all 
stars, without correction. On both diagrams one can see a sequence defined by 
markedly enhanced density of points. Continuous lines in panel (a) represent 
the expected relation between brightness and period for contact binaries 
composed of two equal main sequence stars revolving in orbits with major 
semi-axis  equal to 1.2, 1.35, and 1.5 of their radii, respectively, and 
located at the distance modulus 14.6~mag. The spectral type symbols mark the 
approximate positions of stars of these spectral type on $I_0$ brightness 
axis. 

The existence of the relation in Fig.~10 supports our assumption that most of 
contact binaries from  our catalog are located at the same distance, otherwise 
the points in Fig.~10 would be distributed much more uniformly. The shape of 
the relation is determined by the details of the evolution of contact systems 
and should be compared with expectations of particular theories. In general 
outlines, however, it seems to be consistent  with the following simple 
scheme: the evolutionary not advanced objects of spectral type later than 
about F5 follow the period--brightness relation not very different from that 
for the contact main sequence stars. During their evolution, which in the 
plane of the figure moves a stars from right down to left up, they increased 
their radii not much more than by a factor of 1.5. Stars of earlier spectral 
types increased their radii much more, filling much bigger Roche lobes; they 
are responsible for bending up of the sequence in Fig.~10. The sequence is 
additionally broadened by differences in mass ratio: changing it from 1 
towards smaller values moves the contact binary down, toward shorter periods. 
Thus, the faint stars with periods shorter than about 0.3~day can be 
interpreted as systems with particularly low mass ratio, with at least one 
component being a main sequence star. This scheme fits better the contact 
binaries with equal depths of minima, as can be seen from the Fig.~11: the 
relation period--brightness is much better defined for EWs than for EWa 
objects. 

\Section{ Amplitude Distribution}
Amplitudes of brightness  variations were determined from the decomposition of 
light curves into Fourier harmonics. Fig.~12 shows some examples of the 
observed light curves (empty symbols) and synthetic light curves obtained as 
the sums of five Fourier harmonics (dots), as described in Section~2. 
\begin{figure}[htb]
\vglue-4mm
\centerline{\includegraphics[width=11.8cm]{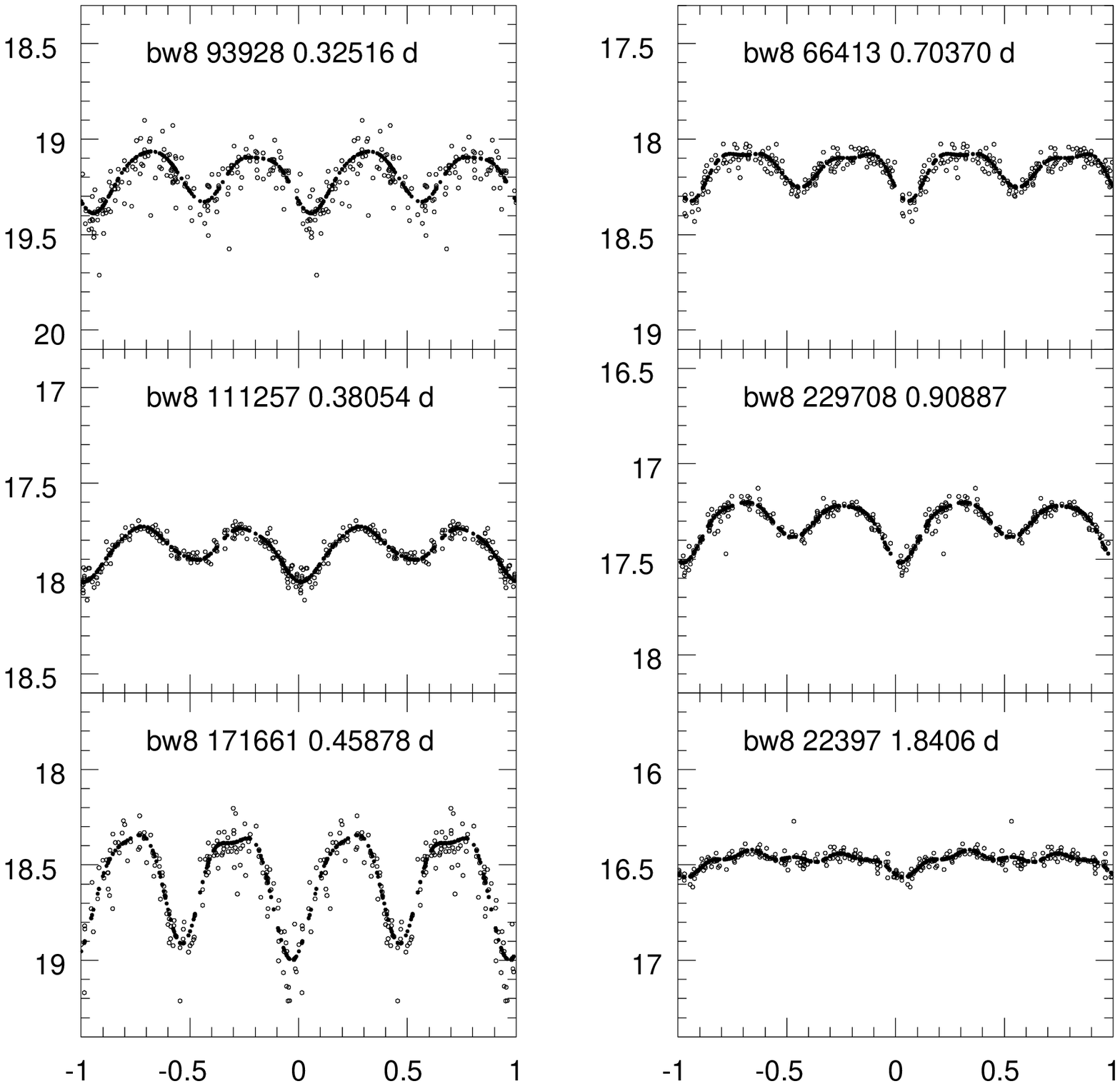}}
\FigCap{Typical observed light curves (empty symbols) and the synthetic light 
curves obtained from the Fourier decomposition (filled symbols). Given are: 
the name of the field, the number in the OGLE database and the period in 
days.} 
\end{figure} 
\begin{figure}[p]
\vglue-7mm
\centerline{\includegraphics[width=9.5cm]{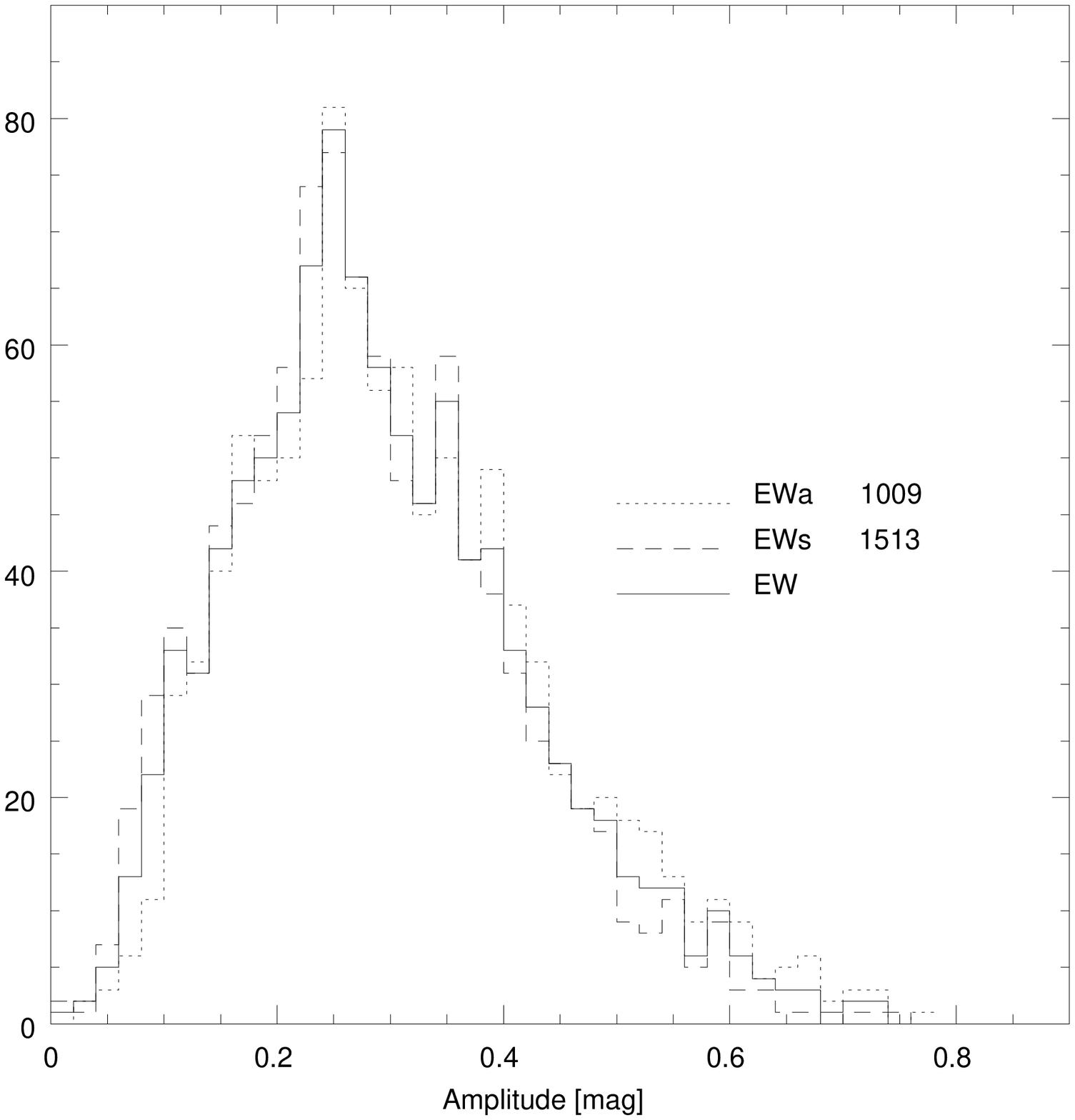}}
\FigCap{Amplitude distributions for the contact binaries with equal and 
different depths of minima. True number of each kind of objects is given.}
\centerline{\includegraphics[width=9.5cm]{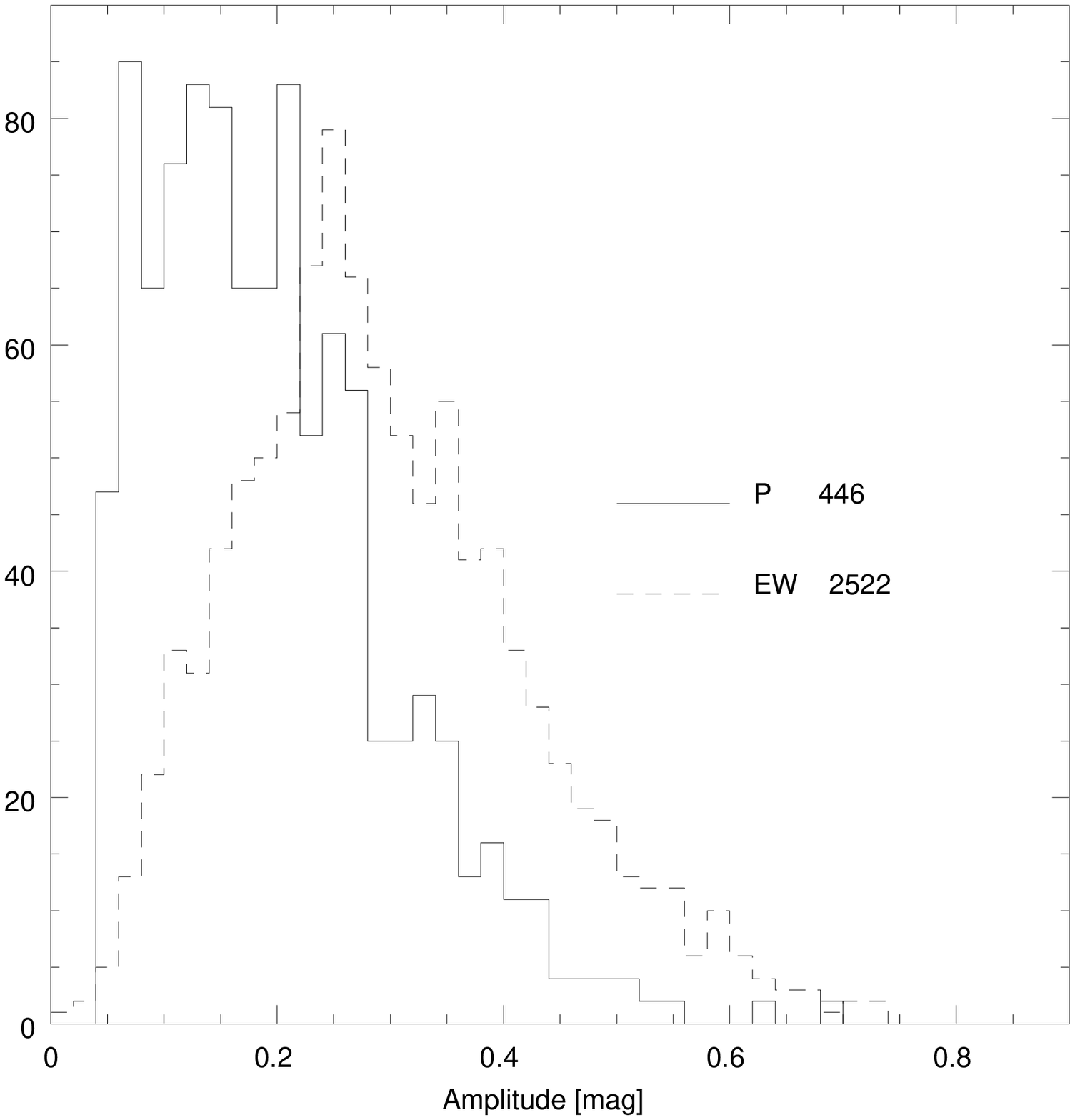}}
\FigCap{Amplitude  distributions for contact binaries (broken line) and for 
pulsating stars (continuous line) from the present catalog. True numbers of 
objects of both kinds are given.} 
\end{figure}

The amplitude of light variation was determined as a difference between the 
brightest and the faintest points in the synthetic light-curve. Because the 
procedure of the Fourier decomposition into five harmonics failed in some  
cases, the number of objects with determined amplitudes is slightly smaller 
(2522) than the number of objects in the catalog. 

Distributions of thus determined amplitudes for EWs and EWa objects are given 
in Fig.~13. As it can easily be seen they are practically identical. This fact 
strongly suggests that -- as it could be expected -- the observed amplitude 
distribution of contact binaries is determined by geometrical factors rather 
than the physical state of the system. 

In Fig.~14  we compare the distribution of amplitudes for all EW stars with 
that for the pulsating stars. These distributions are markedly different, in 
particular at small amplitudes. Amplitude distribution for pulsating stars 
falls down abruptly for values smaller than about 0.05 mag where observational 
errors become comparable with amplitudes themselves. The number of contact 
binaries starts diminishing less steeply and at amplitude value much bigger -- 
about 0.2~mag. This result is rather unexpected. From purely geometrical 
reason the observed number of contact binaries with small amplitudes should 
rise monotonically as the amplitude approaches zero and only obvious selection 
effect makes it fall to zero at zero amplitude. Fig.~14 proves that in the 
case of EW-type light curves this effect starts for some reason at amplitudes 
much bigger than it happens in the case of pulsating variables. 

\Section{Conclusions}
During the first phase of OGLE project we identified 2741 objects with EW-type 
light curves. The majority of them belong most probably to the Galactic bulge. 
If it is true they constitute a sample with homogeneous population 
characteristics. Although all information we have about these stars is limited  
to  the photometric features only, we think that the great number of objects 
in the sample allows some statistical conclusions to be drawn: 

Decomposition of the EW-type light curves into Fourier harmonics provides a 
mean of distinguishing in a statistical way the contact binaries from  
pulsating variable stars. 

There is no statistical difference in distributions of brightness and 
amplitudes of contact binaries with equal and different depths of minima. Both 
groups differ slightly in period distribution. 

The probability of discovery of a  pulsating variable star fainter than ${I= 
18}$~mag is in our sample at most 60\%. Surprisingly, the probability of 
discovery of a faint contact binary is even less, what results in a severe 
selection effect in observed amplitude distribution. 

The character of the statistical relation between period and brightness of 
contact binaries reflects the evolutionary state of their components. If this 
relation refers to objects located in the Galactic bulge then it can be 
treated as an ``isochrone'' of objects that were able to reach the present 
state of contact. 

\Acknow{This work was supported by the KBN Grant 2P03D01418 to M.\ Kubiak}

\end{document}